\documentclass{article}
\usepackage[a4paper,bindingoffset=0.2in,
             left=0.9in,right=0.9in,top=1in,bottom=1in,
             footskip=.25in]{geometry}

\usepackage[utf8]{inputenc}

\usepackage[hang,flushmargin]{footmisc} 
\usepackage{pgfplots} 
\usepackage{tikz}
\pgfplotsset{compat=newest}
\pgfplotsset{plot coordinates/math parser=false}
\pgfplotsset{every tick label/.append style={font=\Large}}
\newlength\figureheight
\newlength\figurewidth
\usetikzlibrary{plotmarks}
\usetikzlibrary{matrix, positioning}
\usetikzlibrary{pgfplots.groupplots}
\usetikzlibrary{decorations.pathmorphing}
\usetikzlibrary{decorations.markings}
\usetikzlibrary{arrows.meta,bending}
\usepackage[font=small,skip=0pt]{caption}
\usepackage[section]{placeins}

\tikzset{
  partial ellipse/.style args={#1:#2:#3}{
    insert path={+ (#1:#3) arc (#1:#2:#3)}
  }
}
\tikzset{%
  >={Latex[width=2mm,length=2mm]},
  base/.style = {rectangle, rounded corners, draw=black, minimum width=4cm, minimum height=1cm, text centered},
  activityStarts/.style = {base, fill=blue!30},
  startstop/.style = {base, fill=blue!30},
  activityRuns/.style = {base, fill=green!30},
  process/.style = {base, minimum width=2.5cm, fill=orange!15},
  contribute/.style = {base, minimum width=2.5cm, fill=yellow!15},
}

\usetikzlibrary{calc,math}
\usetikzlibrary{arrows,matrix,positioning}
\usepackage[normalem]{ulem}
\usepgfplotslibrary{fillbetween}
\usepackage{hyperref}
\usepackage{amssymb}
\usepackage{amsfonts} 
\usepackage{amsmath} 
\usepackage{url}
\usepackage{cite}
\usepackage{amsthm}
\theoremstyle{plain}

\theoremstyle{definition}

\theoremstyle{remark}

\usepackage{algorithm,algcompatible,lipsum}

\usepackage{algpseudocode}
\usepackage{color}
\usepackage{tcolorbox}
\usepackage{nicefrac}
\usepackage{todonotes}

\newcommand{\bs}{\boldsymbol}
\newcommand{\mbf}{\mathbf}
\newcommand{\mbb}{\mathbb}
\newcommand{\mcl}{\mathcal}
 
\usepackage{xstring}
\def\alphabet{abcdefghijklmnopqrstuvwxyzABCDEFGHIJKLMNOPQRST123456789}
\renewcommand{\vec}[1]{
\IfSubStr{\alphabet}{#1}{
\ensuremath{\mathbf{\MakeLowercase{#1}}}
}{
\ensuremath{\boldsymbol{\MakeLowercase{#1}}}
}
}
\newcommand{\mat}[1]{
\IfSubStr{\alphabet}{#1}{
\ensuremath{\mathbf{\MakeUppercase{#1}}}
}{
\ensuremath{\boldsymbol{\MakeUppercase{#1}}}
}
}

\def\R{\mathbb R}
\def\C{\mathbb C}

\newcommand*{\norm}[1]{\left\|#1\right\|}

\newcommand*{\card}[1]{\left|#1\right|}
\def\defeq{:=}

\newcommand*{\Y}[2]{\mathrm{Y}_{#1}^{#2}} 

\usepackage{xcolor}
\newcommand{\ABCor}[1]{\textcolor{teal}{#1}}
\usepackage{acronym}
\acrodef{CS}{Compressed Sensing}
\acrodef{BP}{Basis Pursuit}
\acrodef{OMP}{Orthogonal Matching Pursuit}
\acrodef{AMP}{Approximate Message Passing}
\acrodef{RIP}{Restricted Isometry Property}
\acrodef{BOS}{Bounded Orthonormal System}
\acrodef{IGRF}{International Geomagnetic Reference Field}
\acrodef{SNF}{Spherical Near-Field Antenna Measurements}
\acrodef{SMCs}{Spherical Mode Coefficients}
\acrodef{AUT}{Antenna Under Test}
\acrodef{MIMO}{Multiple Input Multiple Output}
\acrodef{ADMM}{Alternating Direction Method of Multipliers}
\acrodef{ALM}{Augmented Lagrangian Method}
\acrodef{NFFFT}{Near-Field to Far-Field Transformation}
\begin{document}

\title{Optimizing Sensing {\color{black}Matrices} for Spherical Near-Field {\color{black}Antenna} Measurements
}

\author{Arya Bangun\thanks{Forschungszentrum J{\"u}lich,  Germany.}\quad and\quad Cosme Culotta-López\thanks{QuadSAT, Odense,Denmark.}}

\maketitle
\begin{abstract}
In this article, we address the problem of reducing the number of required samples for \ac{SNF} by using \ac{CS}. A condition to ensure the numerical performance of sparse recovery algorithms is the design of a sensing matrix with low mutual coherence. \textcolor{black}{Without fixing any part of the sampling pattern, we propose sampling points that minimize the mutual coherence of the respective sensing matrix by using augmented Lagrangian method.}
Numerical experiments show that the proposed sampling scheme yields a higher recovery success in terms of phase transition diagram when compared to other known sampling patterns, \textcolor{black}{such as} the {spiral and Hammersley sampling schemes}.  Furthermore, we also demonstrate that the application of \ac{CS} with an optimized sensing matrix requires fewer samples than classical approaches {to reconstruct the \ac{SMCs} and far-field pattern.} 
\end{abstract}
 


\section{Introduction}\label{sec1_intro}
Compared to other acquisition methods to characterize an antenna's radiation pattern, \acf{SNF} are a convenient and cost-effective way of ascertaining the radiation characteristics of an \ac{AUT}. {\color{black} Additionally, since the measurement conditions in smaller ranges are easier to control and the obtained results undergo a mathematical transformation that forces them to be physical, their accuracy is higher than for other methods, going as far as being considered the most accurate antenna measurement method \cite{Breinbjerg2016SNF}}. While the biggest challenge of this method may seem the implementation of the mathematical \ac{NFFFT}, one of the most challenging problems is, in fact, the choice of the measurement's sampling strategy. The chosen sampling strategy must ensure that the information of interest is included from an information-theory standpoint and must, also, enable a convenient \ac{NFFFT}. Classically, the Nyquist theorem is applied to the equator of the \ac{AUT} to derive the angular resolution of the sampling scheme, and this angular step is replicated for both classical sampling axes -- azimuth and elevation. The resulting sampling scheme is often called equiangular, and its main drawback is that, considering an information-theory perspective, it results in an equation system with more than double the number of equations, i.e., samples, than variables, i.e., \acf{SMCs} \cite{hansen1988spherical}. As a consequence, one of the strongest objections to this method is its long acquisition time. 

Vibrant research activity has been done to overcome this issue. 
\textcolor{black}{
For instance, the authors of \cite{bucci2001new} use a spiral scanning approach to reduce the density of the samples near the spherical poles, which is a direct implication of using an equiangular sampling.}
Along the same lines, the strategy to reduce sampling points on azimuthal axes has been proposed in \cite{varela2020under}. In addition, the authors also provide an aliasing error estimator to address the aliasing problem that appears naturally when reducing the samples on these axes.
Instead of changing the sampling mechanism, the authors in \cite{foged2016spherical} propose a new \ac{NFFFT} method with downsampling property during the acquisition process. A similar approach has been developed in  \cite{varela2020application} by using non-uniform Fourier transform to accommodate non-equiangular near-field acquisitions, in contrast to the original transformation method that employs conventional Fourier transforms with equiangular sampling points \cite{hansen1988spherical}.

Another strategy to address the long acquisition time is by employing Compressed Sensing (CS), which models the acquisition process as a linear system of equations and applies sparse recovery algorithms, e.g., \ac{BP}, to solve an underdetermined system of equations. This approach stems from the fact that, in most observations, the \ac{SMCs} are sparse or compressible, meaning that only a small amount of coefficients contains high power, as discussed in \cite{cornelius2015investigation, cornelius2016compressed, culotta2017radiationcenter, culotta2021dissertation}. 

\textcolor{black}{
In \ac{CS} theory, if the sensing matrix satisfies certain conditions, for example \ac{RIP}, then it is guaranteed that sparse-recovery algorithms such as \ac{BP} yield a recovery error bounded by the measurement noise and model mismatch \cite{candes2005decoding,candes2006robust}}. 
\textcolor{black}
{Random subgaussian matrices are known to satisfy \ac{RIP} with high probability and thereby have recovery guarantees. This is a considerable roadblock for the practical application of \ac{CS} in \ac{SNF}, since random sampling schemes cannot guarantee in any way that the mechanical movements required to scan them indeed reduce the acquisition time when compared to equiangular sampling schemes \cite{culotta2021dissertation}. Besides, it has been proven in \cite{bandeira2013certifying, tillmann2014computational} that verifying \ac{RIP} property for a given deterministic sensing matrices is an NP-hard problem. Therefore, the \ac{RIP} condition does not provide an adequate guideline for deterministic sensing matrix design}. Apart from \ac{RIP}, another approach to guaranteeing the recovery of sparse signals is the designing low-coherence sensing matrices. Despite providing a weaker reconstruction guarantee when compared to \ac{RIP}, the computation of the mutual coherence is computationally tractable, since it only consists of a dot product of adjacent columns of a sensing matrix. For this reason, designing a low-coherence sensing matrix is preferred in most applications, such as electromagnetic imaging \cite{obermeier2017sensing} and radar \cite{yu2011measurement}. In the mentioned examples, the implementation of \ac{CS} has been conducted by using the mutual coherence as a metric to numerically assess whether that the reconstruction algorithms yield a correct solution with fewer measurement points.

The challenge to construct a low-coherence sensing matrix in \ac{SNF} is that the multipole expansion of electromagnetic fields and antenna measurements requires specific functions, called spherical harmonics and Wigner D-functions, respectively. These functions follow certain polynomial structures, and their optimization is challenging. In this work, we present a method to optimize sensing matrices for \ac{SNF} in terms of their mutual coherence. 
\vspace*{-0.15cm}
\subsection{Related Works}
The construction of a sensing matrix from Wigner D-functions and their specific case, namely spherical harmonics, in the area of \ac{CS} has been  discussed in \cite{rauhut_sparse_2011,burq_weighted_2012,bangun2016sparse,bangun_sensing_2020}. However, these works emphasize more on the design of \ac{RIP} sensing matrices with random sampling and theoretical recovery guarantee using \ac{BP}. Although we have a bound on the minimum required number of measurements, in practice, this bound seems pessimistic, as discussed in \cite{hofmann2019minimum}, since the \ac{BP} algorithm is still able to numerically reconstruct the sparse coefficients even with fewer samples distributed in a spiral pattern. This investigation leads to the question of whether other types of deterministic sampling points can be designed to yield a higher recovery success, even if just numerically.

In \cite{bangun_coherence_2018,bangun_sensing_2020}, the authors aim to show that a specific sampling pattern, e.g., the equiangular sampling scheme, leads to the construction of a high coherence sensing matrix and fails to reconstruct the sparse signal which, in turn, excludes a construction of sensing matrices based on such a sampling strategy for \ac{CS} applications. Additionally, the authors derive an achievable coherence bound that can be used to design a low coherence sensing matrix for \ac{SNF} setting \cite{culotta2018compressedsampling}. 
However, the drawback to this approach is that, while fixing a specific pattern on the elevation axes is advantageous for mechanical scanning systems, some degree of freedoms are lost in comparison to a full optimization on the whole spherical surface. 

\vspace*{-0.15cm}
\subsection{Summary of Contributions}
The main contributions of our paper are as follows:
\begin{itemize}
\item We propose a gradient-based method by smoothing the objective function to minimize the coherence in terms of $\ell_p$-norm. Thereby, we only have to set a large enough $p$ to approximate the maximum norm. 
\item Besides, we also propose an \ac{ALM}-based proximal method. This approach provides an efficient strategy to decompose the original problem into subproblems. By using the proximal for the $\ell_{\infty}$-norm, we do not have to directly compute gradient or sub-gradient $\ell_{\infty}$-norm. 
\item We also provide numerical simulations to evaluate the coherence of a sensing matrix from both algorithms and compare them to the coherence of a sensing matrix constructed from well-known sampling patterns. 
We  numerically show that our proposed sampling scheme outperform other well-known sampling patterns and deliver results similar to their random counterparts. 
\item At last, we apply optimized sensing matrices for experimental data and perform sparse \ac{SMCs} reconstruction, and use these to reconstruct the far-field pattern. Numerically, we show that by using optimized sensing matrices for \ac{CS} it is possible to reduce the number of samples to reconstruct an \ac{AUT}'s far-field pattern.
\end{itemize}

\section{Compressed Sensing}\label{sec2_cs}

In most signal acquisition processes, it is enough to model the {measurement processes} with a linear system as follows:
\begin{equation}
\label{eq:linear_meas}
    \mbf y = \mbf A \mbf x,
\end{equation}
where $\mbf A \in \C^{K \times L}$ is the sensing matrix. \ac{CS} provides a framework to recover the sparse structure of $\mbf{x}$ from a smaller dimension of the acquired signal $\mbf y \in \C^K$, i.e., $K < L$. {\acf{BP} is one of the well-known algorithms to estimate the sparse vector $\mbf{x}$ by solving the following optimization problem}
\begin{equation}
\tag{P1}\label{ch2:eq P1}
\begin{aligned}
& \underset{\mbf{\hat x}}{\text{minimize}}
& & \norm{\mathbf{\hat x}}_1
& \text{subject to}
& & \mbf y = \mbf A \mbf{\hat x}.
\end{aligned}
\end{equation}
Theoretically speaking, it is necessary to have a fundamental limit on the required number of measurements $K$ to ensure the unique reconstruction of sparse vector $\mbf x$, meaning that the estimated vector $\mbf{\hat x}$ and the ground truth vector $\mbf{x}$ are close enough in terms of small numerical differences. Low coherence of the matrix $\mbf A$ are well-known properties to provide a sufficient condition and robustness of the reconstruction of sparse vector $\mbf{x}$ \cite{gribonval2003sparse,candes2005decoding,candes2006robust}.

For a matrix $\mbf A=[\vec{a}_1 \dots \vec{a}_L]\in\mbb C^{K \times L}$, the {\color{black}mutual} coherence is defined as
\[
 \mu
 (\mbf{A})\defeq\max_{1\leq i<j\leq L}\frac{\card{\langle\vec{a}_i,\vec{a}_j\rangle}}{\norm{\vec{a}_i}_2  \norm{\vec{a}_j}_2},
\]
where the lowest coherence we can get is lower bounded by the Welch bound \cite{welch_lower_1974}, which is given as
$$
\mu(\mbf A) \geq \sqrt{\frac{L - K }{K(L-1)}}.
$$
Nonetheless, constructing a low coherence sensing matrix is generally considered important for practical purposes. Before discussing the construction of the sensing matrix for \ac{SNF}, the next section will cover the measurement setting {\color{black}and the mathematical framework of} \ac{SNF}.
\section{Spherical Near-Field Measurements}\label{sec3_snf}
{\color{black}Electromagnetic fields can be modelled as} a superposition of harmonic functions and have, therefore, a periodic structure. Specifically, the radiated fields that are acquired by a probe antenna can be expanded by orthogonal basis functions that solve Maxwell's equations \cite[Eq 4.43]{hansen1988spherical}. We can express the relation as follows:
\begin{equation}
\footnotesize
w(r,\theta,\phi,\chi)= \sum \limits_{\mu=-v_\text{max}}^{v_\text{max}}\sum\limits_{s=1}^2\sum\limits_{n=1}^N\sum\limits_{m=-n}^n \mathrm{D}_{\mu,m}^n(\theta,\phi,\chi)  uT_{smn} P_{s\mu n} (kr),
\label{sec3_trans_eq}
\end{equation} 
where $\mathrm{D}_{\mu,m}^n(\theta,\phi,\chi) := e^{jm\phi} \mathrm{d}_{\mu,m}^n(\cos \theta) e^{j\mu\chi} $ are the Wigner D-functions of degree $n$ and orders $\mu,m$. Mathematically, these functions act as basis functions in the three-dimensional rotation space to represent electromagnetic fields. The function $\mathrm{d}_{\mu,m}^n(\cos \theta)$ represents the Wigner d-functions, defined by
\begin{equation} \label{Wigner_d}
\mathrm{d}_{\mu,m}^{n}(\cos \theta)\defeq \omega \sqrt{\gamma} \sin^{\xi} \bigg(\frac{\theta}{2}\bigg)\cos^{\lambda}\bigg(\frac{\theta}{2}\bigg) P_{\alpha}^{(\xi,\lambda)}(\cos \theta),
\end{equation}
where $\gamma = \frac{\alpha!(\alpha + \xi + \lambda)!}{(\alpha+\xi)!(\alpha+\lambda)!}$, $\xi=\card{m - \mu}$, $\lambda=\card{m + \mu}$, $\alpha= n -\big(\frac{\xi+\lambda}{2}\big)$. For $\mu \geq m $ the value $ \omega = 1$ otherwise $(-1)^{\mu-m}$.
The function $P_{\alpha}^{(\xi,\lambda)}$ represents the Jacobi polynomials. The Wigner D-functions bear a strong resemblance with a set of orthogonal basis functions on the sphere: the spherical harmonics. The relation between these basis functions can be derived by considering the order $\mu = 0$:
\begin{equation}
\mathrm{D}_{0,m}^n(\theta,\phi,\chi) =  (-1)^m \sqrt{\frac{4\pi}{2n + 1}} \Y{n}{m}(\theta, \phi),
\label{eq.sh_wign}
\end{equation}
where the spherical harmonics $\Y{n}{m}(\theta, \phi)$ can be expressed as the product of associated Legendre and trigonometric polynomials
$$\Y{n}{m}(\theta,\phi) = \sqrt{\frac{2n + 1}{4\pi}\frac{(n-m)!}{(n + m )!}} P_n^m(\cos \theta) e^{im\theta}.$$

\begin{figure}[!htb] 
\centering
     \scalebox{0.6}{
\begin{tikzpicture}[scale=0.7]
		\tikzmath{\xLength = 5;
							\yLength = 5;
							\zAngle = 60;
							\radius = 3;
							\xLengthProbe = \xLength/2;
							\yLengthProbe = \yLength/2;
							\radiusProbe = \radius/2.5;
							\xAUT = 4;
							\yAUT = 2;
							\zAUT = 2;
							\orientation = -60;
							\angleCenter = 90+atan((\yAUT-\zAUT*cos(\zAngle))/(\xAUT-\zAUT*sin(\zAngle)))-180*sign((sign(\xAUT-\zAUT*sin(\zAngle))-1));
							\minSphere = sqrt(\xAUT^2+\yAUT^2+(\zAUT*sin(\zAngle)^2));}
		\begin{scope}
				\node (oriCenter) at (0,0) {};
				\node[label={[black]below left:$y$}] (oriY) at (\xLength, 0) {};
				\node[label={[black]below right:$z$}] (oriZ) at (0,\yLength) {};
				\node[label={[black]right:$x$}] (oriX) at (-{\xLength*cos(\zAngle)} ,{-\yLength*sin(\zAngle)} ) {};
				\draw[black,thick,-latex'] (oriCenter.center) -- (oriX.center);
				\draw[black,thick,-latex'] (oriCenter.center) -- (oriY.center);
				\draw[black,thick,-latex'] (oriCenter.center) -- (oriZ.center);
				\draw[thick] (oriCenter) circle (\radius);
				\draw[black,thick,-latex'] (oriCenter.center) [partial ellipse=90:0:{\radius/2} and {\radius/2}];
				\node[label={[black]above right:$\theta$}] (theta) at ({1*\radius/7},{2*\radius/7+0.3}) {};
				\draw[thick,dashed] (oriCenter.center) [partial ellipse=0:180:{\radius} and 1.4];
				\draw[thick] (oriCenter.center) [partial ellipse=180:360:{\radius} and 1.4];
				\draw[thick,dashed] (oriCenter.center) [partial ellipse=90:270:2 and \radius];
				\draw[thick] (oriCenter.center) [partial ellipse=0:90:2 and \radius];
				\draw[thick] (oriCenter.center) [partial ellipse=270:360:2 and \radius];
				\node (intersection) at (1.89,-1.13){};
				\draw[black,thick] (oriCenter.center) -- (intersection.center);
				\draw[black,thick,-latex'] (oriCenter.center) [partial ellipse=240:325:{\radius/6} and {\radius/6}];
				\node[label={[black]below:$\phi$}] (phi) at ({0},{-\radius/12}) {};
		\end{scope}
		\begin{scope}[shift = {{({10} ,{0} )}}, rotate = \orientation]
			\node (oriCenterProbe) at (0,0) {};
			\node[label={[black]below left:$x'$}] (oriXProbe) at (\xLengthProbe, 0) {};
			\node[label={[black]below right:$y'$}] (oriYProbe) at (0,\yLengthProbe) {};
			\node[label={[black]below right:$z'$}] (oriZProbe) at (-{\xLengthProbe*cos(\zAngle)} ,{-\yLengthProbe*sin(\zAngle)} ) {};
			\draw[black,thick,-latex'] (oriCenterProbe.center) -- (oriXProbe.center);
			\draw[black,thick,-latex'] (oriCenterProbe.center) -- (oriYProbe.center);
			\draw[black,thick,-latex'] (oriCenterProbe.center) -- (oriZProbe.center);
			\draw[thick] (oriCenterProbe) circle (\radiusProbe);
			\draw[thick,dashed] (oriCenterProbe.center) [partial ellipse=0:180:{\radiusProbe} and 1];
			\draw[thick] (oriCenterProbe.center) [partial ellipse=180:360:{\radiusProbe} and 1];
			\draw[thick] (oriCenterProbe.center) [partial ellipse=90:270:0.8 and \radiusProbe];
			\draw[thick,dashed] (oriCenterProbe.center) [partial ellipse=0:90:0.8 and \radiusProbe];
			\draw[thick,dashed] (oriCenterProbe.center) [partial ellipse=270:360:0.8 and \radiusProbe];
		\end{scope}
		\draw[black,thick] (oriCenter.center) -- (oriCenterProbe.center);
		\node[label={[black]below: $r$}] (oriCenter) at ({5.5, 1.1}){};
		\begin{scope}[shift = {{({7.2} ,{0} )}}, rotate = 0]
			\node (chiCenter) at (0,0) {};
			\draw[black,thick,-latex'] (chiCenter.center) [partial ellipse=15:345:0.35 and 1.2];
			\node[label={[black]below left:$\chi$}] (chi) at (0,0) {};
		\end{scope}
\end{tikzpicture}}  

\caption{Measurement settings for \ac{SNF}}
\label{fig:SNF_Setting}
\end{figure}
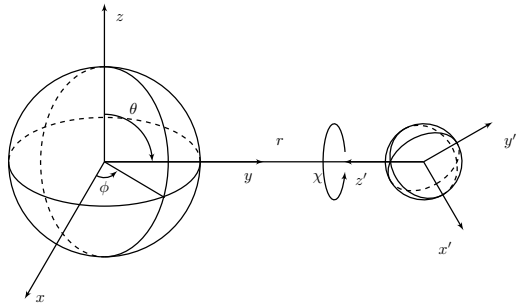

Classically, it is considered that, after a certain degree $n = N$, the contribution of higher modes to the total power of the expansion is limited. The truncation constant applied is typically calculated as $N = k r_{min} + N_0$, where $k=\nicefrac{2\pi}{\lambda}$ is the wavenumber, $r_{min}$ is the minimum radius of the sphere that encloses the \ac{AUT}, and $N_0$ is a constant for accuracy, whereby $N_0 = 10$ is supported in the literature \cite{hansen1988spherical}. Additionally, {\color{black}$v_\text{max}$ depends on the distance the probe antenna is being considered at, i.e., $v_\text{max} = N$ generally, since its modes are displaced with respect to the center of the coordinate system of the measurement.} 
In \eqref{sec3_trans_eq}, $u$ is the input signal to the \ac{AUT} and $w(r, \theta, \phi, \chi)$ is the radiated field acquired by a given measurement probe at a distance $r$, with elevation $\theta \in [0,\pi]$ and with azimuth angle $\phi \in [0, 2\pi)$. The response of the probe itself, i.e., the radiation pattern of the probe itself, affects the measured signal as well. The probe response constants $P_{s\mu n} (kr)$ describe this behavior in the acquisition process. Additionally, another factor comes into play: the polarization characteristics of both \ac{AUT} and probe. To measure it, the polarization angle $\chi \in [0, 2\pi)$ is added. Fig.~\ref{fig:SNF_Setting} shows the described measurement geometry.
The transmission coefficients $T_{smn}$ can be decomposed into transverse electric and transverse magnetic coefficients, described by $T_{1mn}$ and $T_{2mn}$, respectively, and are useful to determine properties of the \ac{AUT} in the far-field region. The objective of \ac{SNF} measurements is, ultimately, to solve the inverse problem that derives the transmission coefficients $T_{smn}$ from near-field measurements by solving the linear equations resulting from \eqref{sec3_trans_eq}. Once the $T_{smn}$ are known, the \ac{AUT}'s far-field characteristic can be easily computed by solving \eqref{sec3_trans_eq} for $r\to\infty$, which represents the theoretical far field.
\vspace*{-0.15cm}
\subsection{General Case} \label{Sec:General}
Suppose we acquire $K$ samples of radiated fields on the unit sphere $w(\theta_i,\phi_i. \chi_i)$ for $i \in [K]$. {We consider the general case with $v_\text{max} = n$}. 
Thereby, we can rewrite \eqref{sec3_trans_eq} into
\begin{equation}
\begin{aligned}
 \sum\limits_{n = 1}^N \sum\limits_{m = -n}^n \sum\limits_{\mu = -n}^n \mathrm{D}_{\mu m}^n(\theta_i,\phi_i,\chi_i)\left(c_{1mn\mu}  + c_{2mn\mu} \right),
\label{sec3_expand}
\end{aligned}
\end{equation}
where we write $c_{smn\mu}= uT_{snm}P_{s\mu n}(k)$ as the product between transmission coefficients and the probe response constants with input signal $u$. We can then represent equation \eqref{sec3_expand} in matrix and vector form as in \eqref{eq:linear_meas},
where the elements of vector $\mbf{y} \in \C^{K}$ and coefficients $\mbf x \in \C^{L}$ are given by $y_i = w(\theta_i,\phi_i,\chi_i)\quad \text{for}\quad i \in [K]$ and $x_q~=~\left(c_{1 m(q) n(q) \mu(q)}  + c_{2 m(q) n(q) \mu(q)} \right) \quad \text{for}\quad q \in [L]$,
respectively. The sensing matrix $\mat{A} \in \C^{K \times L}$ is given by 
\begin{equation}
\small
\mat{A}=
\begin{pmatrix}
  \mathrm{D}^{1}_{-1,-1}(\theta_1,\phi_1,\chi_1)&\dots& \mathrm{D}^{B-1}_{B-1,B-1}(\theta_1,\phi_1,\chi_1) \\
  \vdots\\
  \mathrm{D}^{1}_{-1,-1}(\theta_K,\phi_K,\chi_K)&\dots& \mathrm{D}^{B-1}_{B-1,B-1}(\theta_K,\phi_K,\chi_K)
\end{pmatrix},
\label{sec_3.sensingmatrix}
\end{equation}
where columns of this matrix consist of $K$ different samples of Wigner D-functions. For clarity, we can write the elements of a matrix $\mat{A} \in \C^{K \times L}$  as
\begin{equation}
A_{iq} = \mathrm{D}_{\mu(q) m(q)}^{n(q)}(\theta_i,\phi_i,\chi_i),
\end{equation}
where the column dimension $L = \frac{4N^3 + 12N^2 + 11N}{3}$, which is determined by a combination of degree and orders $n(q),m(q),\mu(q)$, respectively.  
\vspace*{-0.15cm}
\subsection{Specific Case}\label{Sec:Specific}   
As discussed in \cite{hansen1988spherical, wacker1975non}, assuming only the $\mu = \pm 1$ modes of the probe contribute significantly to the measurement is enough for most cases and considerably reduces the computational complexity of the problem. This assumption is valid for most linearly polarized probes. 
Elaborating further, and as discussed in \cite[Eq 4.99]{hansen1988spherical}, a linearly polarized probe has symmetric and anti-symmetric properties for each transverse electric $s = 1$ and magnetic $s = 2$ mode, as expressed by $ c_{smn(\mu=-1)} =  c_{smn(\mu=1)}
$ for $s = 1$ and $c_{smn(\mu=-1)} =  -c_{smn(\mu=1)} $ for $s = 2$.
Therefore, we can treat 
\eqref{sec3_expand}
 as the superposition of two expansions and rewrite it into
\begin{equation}
\begin{aligned}
&\sum\limits_{n = 1}^N \sum\limits_{m = -n}^n \bigg(\mathrm{D}_{1 m}^n(\theta_i,\phi_i,\chi_i) +\mathrm{D}^n_{-1 m}(\theta_i,\phi_i,\chi_i) \bigg) c_{1mn1}  + \\
&\sum\limits_{n = 1}^N\sum\limits_{m = -n}^n \bigg(\mathrm{D}_{1 m}^n(\theta_i,\phi_i,\chi_i) -  \mathrm{D}^n_{-1 m}(\theta_i,\phi_i,\chi_i) \bigg) c_{2mn1}
\label{sec3_expand2}
\end{aligned}
\end{equation}
or, equivalently, in matrix form as 
\begin{equation}
\begin{aligned}
\mathbf{y} = \left[\begin{array}{c} \mathbf{A_1} \quad \mathbf{A_2} \end{array}\right]\left[\begin{array}{c} \mathbf{x_1}\\ \mathbf{x_2}  \end{array}\right], 
\end{aligned}
\end{equation}
where the elements of matrices $\mbf{A_1},\mbf{A_2} \in \C^{K \times L}$ are, in turn, given by
\begin{equation}
A_{iq}^1 =  \mathrm{D}_{1 m(q)}^{n(q)}(\theta_i,\phi_i,\chi_i) + \mathrm{D}_{-1 m(q)}^{n(q)}(\theta_i,\phi_i,\chi_i),
\end{equation}
\vspace*{-0.5cm}
\begin{equation}
A_{iq}^2 = \mathrm{D}_{\mu(q) m(q)}^{n(q)}(\theta_i,\phi_i,\chi_i) - \mathrm{D}_{\mu(q) m(q)}^{n(q)}(\theta_i,\phi_i,\chi_i),
\end{equation}
and for each matrix we have column dimension $L$. Similarly, the transmission coefficients can be expressed as element vectors $\mbf{x}_1,\mbf{x}_2 \in \C^L$ and are given by $x^1_{q} = c_{1m(q)n(q)1}$ and $x^2_{q} = c_{2m(q)n(q)1}$ for a combination degree $m(q)$ and order $n(q)$, respectively. 

\section{Optimization of Sensing Matrices} \label{sec4_opt}
{Since the structure of a sensing matrix is restricted by certain properties of Wigner D-functions, the coherence optimization, therefore, boils down to finding optimal sampling points that yield a low coherence sensing matrix constructed from sampled Wigner D-functions.} The coherence of a sensing matrix from Wigner D-functions is given by 
\begin{equation} \label{coherence_Wigner}
\begin{aligned}
\mu  &= \underset{ 1 \leq r < q \leq L}{\text{max}} \card{g_{q,r} \left(\bs \theta,\bs \phi,\bs \chi \right)}\\   
\end{aligned},
\end{equation}
where $\card{g_{q,r} \left(\bs \theta,\bs \phi,\bs \chi \right)}$ is expressed as
\begin{equation}
    \small
 \card{\sum_{i = 1}^{K}\frac{\mathrm{D}^{n{(q)}}_{\mu{(q)},m{(q)}}(\theta_i,\phi_i,\chi_i)  \overline{\mathrm{D}^{n{(r)}}_{\mu{(r)},m{(r)}}(\theta_i,\phi_i,\chi_i)}}{\norm{\mathrm{D}^{n{(q)}}_{\mu{(q)},m{(q)}}(\boldsymbol\theta,\boldsymbol\phi,\boldsymbol\chi)}_2 \norm{\mathrm{D}^{n{(r)}}_{\mu{(r)},m{(r)}}(\boldsymbol\theta,\boldsymbol\phi,\boldsymbol\chi)}_2}}
 \label{eq:product}
\end{equation}
and the vector $\mathrm{D}^{n}_{\mu,m}(\boldsymbol\theta,\boldsymbol\phi,\boldsymbol\chi)$ is, in turn, 
\[
\small \mathrm{D}^{n}_{\mu,m}(\boldsymbol\theta,\boldsymbol\phi,\boldsymbol\chi)\defeq
\begin{pmatrix}
 \mathrm{D}^{n}_{\mu,m}(\theta_1,\phi_1,\chi_1),
 \hdots,
  \mathrm{D}^{n}_{\mu,m}(\theta_K,\phi_K,\chi_K)
\end{pmatrix}^T.
\]
The problem of optimizing the coherence of a sensing matrix from Wigner D-functions can be seen as finding the distribution of sampling points on $\theta \in [0, \pi]$ and $\phi, \chi \in [0, 2\pi)$ that minimizes \eqref{coherence_Wigner} as follows:

\begin{equation}
\begin{aligned}
\centering
& \underset{\bs \theta, \bs \phi, \bs \chi \in \R^K}{\text{min}}
& & \underset{ 1 \leq r < q \leq N}{\text{max}} \card{g_{q,r}(\bs \theta,\bs \phi,\bs \chi)}\\
\end{aligned}\quad.
\label{eq:Opt_General}
\end{equation}

This problem equals to solving the min-max optimization of a non-convex objective function.  In this work, we propose two approaches to tackle this optimization problem.

\subsection{Gradient-based $\ell_p$-norm}
The non-smoothness of the absolute value function poses a challenge in directly minimizing the coherence. However, 
as presented in the notation in Section \ref{Sect:Appendix}, the $\ell_{\infty}$-norm can be approximated by the $\ell_p$-norm with a large enough $p$, as in
\begin{equation}
\begin{aligned}
\centering
&\underset{\bs \theta, \bs \phi, \bs \chi \in \R^K}{\text{min}}
& &\underset{p \rightarrow \infty}{\lim} \left(\underset{1 \leq r < q \leq N}{\sum} \card{g_{q,r}(\bs \theta,\bs \phi,\bs \chi)}^{p}\right)^{1/p}\\
\end{aligned}\;\;.
\label{eq:Opt_Approx}
\end{equation}

In contrast to the strategy in \cite{bangun2021tight} where only to optimize the azimuth and polarization angles $\phi,\chi \in [0,2\pi)$ to achieve the lower bound derived from equispaced sampling points on $\theta \in [0,\pi]$, the challenge when optimizing $\theta$, as opposed to variables $\phi$ and $\chi$, we have a more complex structure in terms of trigonometric functions and Jacobi polynomials, as expressed in \eqref{Wigner_d}. We first show the derivative with respect to $\theta$ as follows
\begin{equation} 
\footnotesize
\label{sec4_deriv}
\begin{aligned}
\nabla_{\bs \theta} &\left(\underset{ 1 \leq r < q \leq L}{\sum} \card{{g_{q,r}(\bs \theta,\bs \phi,\bs \chi)}}^{p}\right)^{\frac{1}{p}} =\left(\underset{ 1 \leq r < q \leq L}{\sum} \card{{g_{q,r}(\bs \theta,\bs \phi,\bs \chi)}}^{p}\right)^{\frac{1 - p}{p}} \\& \times\left(\underset{ 1 \leq r < q \leq L}{\sum} \card{{g_{q,r}(\bs \theta,\bs \phi,\bs \chi)}}^{p - 1}\nabla_{\bs \theta} \card{g_{q,r}(\bs \theta,\bs \phi,\bs \chi)} \right),\\
\end{aligned}
\end{equation} 
where $\nabla_{\bs \theta} \card{g_{q,r}(\bs \theta,\bs \phi,\bs \chi)}$ can be derived directly by taking the expression in \eqref{eq:product}, where the Wigner D-function is given by $\mathrm{D}_{\mu,m}^n(\theta,\phi,\chi) := e^{jm\phi} \mathrm{d}_{\mu,m}^n(\cos \theta) e^{j\mu\chi} $. The derivative with respect to $\bs \theta$ can be done by applying the product rule of derivatives, so that the derivative of Wigner d-functions, $\frac{\mathrm{d}}{\mathrm{d} \theta}\left(\mathrm{d}_{n}^{\mu,m}(\cos \theta)\right) $, can be expressed as
\begin{equation*}
\begin{aligned}
\small
 &\left({\frac{\xi \sin \theta}{2(1 - \cos \theta)}} - {\frac{\lambda \sin \theta}{2(1 + \cos \theta)}}  \right) \mathrm{d}_{n}^{\mu,m}(\cos \theta)  - {N_l}\sin \theta\,\omega \sqrt{\gamma}\\&\times \left(\frac{\xi + \lambda + \alpha + 1}{2}\right) \sin^{\xi}\left(\frac{\theta}{2} \right) \cos^{\lambda} \left( \frac{\theta}{2}\right)\times  P_{\alpha-1}^{\xi+1,\lambda + 1}(\cos \theta),
\end{aligned}
\end{equation*}
where the definition of $\mathrm{d}^{n}_{\mu,m}(\cos \theta)$  is given in \eqref{Wigner_d} and the chain rule is subsequently applied with respect to $\theta$, i.e., $\frac{\mathrm{d}}{\mathrm{d}\theta} f(\cos \theta) = -\sin \theta \frac{d}{d \cos \theta} f(\cos \theta) $. In addition, the $k$-th derivative of the Jacobi polynomials is given by
$$\small \frac{\mathrm{d}^k}{\mathrm{d} \cos^k \theta} { P_{\alpha}
^{\xi , \lambda} (\cos \theta)} = \frac{\Gamma(\xi + \lambda + \alpha + 1 + k)}{2^k \Gamma(\xi  + \lambda + \alpha + 1)} P_{\alpha - k}
^{\xi + k, \lambda + k} (\cos \theta)$$
where $\Gamma(x) = (x-1)!$. 

In addition, $\nabla_{\bs \phi}   \card{g_{q,r}(\bs \theta,\bs \phi,\bs \chi)}$ and $\nabla_{\bs \chi}   \card{g_{q,r}(\bs \theta,\bs \phi,\bs \chi)}$ are straightforward 
because we have trigonometric polynomials for orders $\mu, m$. Since the coherence should be evaluated in terms of the product of adjacent columns, the next step is to perform the product rule of derivative to complete the calculation of the gradient.

{Thereby, the variables $\bs \theta, \bs \phi$, and $\bs \chi$ can be iteratively updated by incorporating the gradient. For simplicity, we show the update for $\bs \theta$ at the $i$-th iteration as follows
\begin{equation}\label{eq:update_gd_theta}
\small
\bs{\theta}^{(i)} = \bs{\theta}^{(i-1)} - \eta \nabla_{\bs \theta} \left(\underset{ 1 \leq r < q \leq L}{\sum} \card{{g_{q,r}(\bs \theta^{(i-1)},\bs \phi^{(i-1)},\bs \chi^{(i-1)})}}^{p}\right)^{1/p} ,
\end{equation}
where the variable $\eta$ is the step size parameter. The same approach holds for $\bs \phi$ and $\bs \chi$. The summary of the procedure is given in Algorithm \ref{algo_gd}.}

\begin{algorithm}
    \caption{Gradient Descent of $\ell_p$-norm}\label{algo_gd}
    \begin{algorithmic}[1]
     
        \State \textbf{Initialization:} 
		\begin{itemize}
		\item Initial angles $\bs \theta^{(0)}, \bs \phi^{(0)}, \bs \chi^{(0)} \in \R^K$ uniformly random on interval $[0,\pi]$ and $[0,2\pi)$
		\item Step size $\eta \in \R$ and maximum iteration $T$
		\item Large enough $p$ for $\ell_{p}$-norm, initial coherence $\rho_0 = 1$.
 
		\end{itemize}
    \For{each iteration $i \in [T]$}  
    \State Update  $\bs \theta^{(i)}, \bs \phi^{(i)}$, and $\bs \chi^{(i)}$ as in \eqref{eq:update_gd_theta} 
    \If {$\underset{ 1 \leq r < q \leq N}{\text{max}} \card{g_{q,r}(\bs \theta^{(i)},\bs \phi^{(i)},\bs \chi^{(i)})} < \rho_{i-1}$ }
    	\State{$\rho_i = \underset{ 1 \leq r < q \leq N}{\text{max}} \card{g_{q,r}(\bs \theta^{(i)},\bs \phi^{(i)},\bs \chi^{(i)})}$}
   	\Else 
   		\State{$\rho_{i} = \rho_{i-1}$}
	\EndIf
    \EndFor   
    \end{algorithmic}
  \end{algorithm} 

\subsection{\ac{ALM}-based algorithm} 

Instead of smoothing the $\ell_{\infty}$-norm through the $\ell_p$-norm for $p \in [2,\infty)$, we propose another approach by using the proximal method. In this case, we are not required to compute the gradient or subgradient directly. In the case of $\ell_{\infty}$-norm, its proximal operator can be written as the projection onto its dual norm, which is the $\ell_1$-norm \cite{parikh2014proximal}. Suppose we have a vector $\mbf x \in \R^L$, then the proximal operator of $\norm{\mbf x}_{\infty}$ is given by
\[
\text{prox}_{\norm{.}_\infty}\left(\mbf x\right) = \mbf x - \text{Proj}_{\norm{.}_1}\left( \mbf x \right).
\]
The projection onto the $\ell_1$-norm is well-studied, for instance, in \cite{duchi2008efficient} and can be directly implemented. 

The proximal method can be combined with the \ac{ALM} method to solve the optimization problem for minimizing the coherence of a sensing matrix.
Let us assume we have set $\mcl{S} = \{\left(r,q\right) | 1 \leq r < q \leq L\}$, which represents the combinations of adjacent column matrices to evaluate the coherence with cardinality $\card{\mcl{S}} = J$. 
Thereby, the optimization \eqref{eq:Opt_General} can be written as
\begin{equation}
\begin{aligned}
\centering
& \underset{\mbf{z} \in \C^J\atop\bs \theta, \bs \phi, \bs \chi \in \R^K}{\text{min}} \norm{\mbf z}_{\infty}
\text{subject to}
& & z_j = g_{q(j),r(j)}(\bs \theta,\bs \phi,\bs \chi), \quad {j \in [J]}
\end{aligned}
\label{eq:Opt_ALM}
\end{equation}
We can write the scaled form of the augmented Lagrangian of the problem \eqref{eq:Opt_ALM} as 
$$
\small
\mcl{L}\left(\bs \theta, \bs \phi, \bs \chi, \mbf z, \mbf u \right) = \norm{\mbf z}_{\infty} + f\left(\bs \theta,\bs \phi,\bs \chi, \mbf z, \mbf u \ \right),
$$
where the constraint can be written as $f\left(\bs \theta,\bs \phi,\bs \chi, \mbf z, \mbf u\right) =  \sum_{j = 1}^J \frac{\tau}{2} \card{z_j - g_{q(j),r(j)}(\bs \theta,\bs \phi,\bs \chi) + u_j}^2$.

Therefore, the procedure to optimize all variables in the augmented Lagrangian $\mcl{L}\left(\bs \theta, \bs \phi, \bs \chi, \mbf z, \mbf u \right)$ is now written as follows
\begin{align}
   & \mbf z^{\left(i \right)}= \underset{\mbf z}{\text{arg min}} \quad \mcl{L}\left(\bs \theta^{\left(i -1 \right)}, \bs \phi^{\left(i -1 \right)}, \bs \chi^{\left(i -1 \right)}, \mbf z, \mbf u^{\left(i -1 \right)} \right) \label{eq:update_z}\\
    &\bs \theta^{\left(i \right)},\bs \phi^{\left(i \right)},\bs \chi^{\left(i \right)}= \underset{\bs \theta,\bs \phi, \bs \chi}{\text{arg min}}  \quad \mcl{L}\left(\bs \theta , \bs \phi , \bs \chi , \mbf z^{\left(i -1 \right)}, \mbf u^{\left(i -1 \right)} \right)\label{eq:update_ang}\\
    &\small{\mbf u^{\left(i\right)} = \mbf u^{\left(i - 1\right)} + \tau \left(\mbf z^{\left(i\right) } - g_{q(j),r(j)}(\bs \theta^{\left(i \right)},\bs \phi^{\left(i \right)},\bs \chi^{\left(i \right)}) + \mbf u^{\left(i - 1\right)} \right)} \label{eq:update_u}
\end{align}
 
The complete \ac{ALM} algorithm can now be expressed as in Algorithm \ref{algo_ALM}.
\begin{algorithm} 
    \caption{\ac{ALM}-based algorithms}\label{algo_ALM}
    \begin{algorithmic}[1]
       
      \State \textbf{Initialization:} 
\begin{itemize}
    \item Initial angles $\bs \theta^{\left(0\right)} \bs \phi^{\left(0\right)}, \bs \chi^{\left(0\right)} \in \R^K$ uniformly random $[0,\pi]$ and $[0,2\pi)$
    \item Initial auxiliary variables $\mbf z^{\left(0\right)}, \mbf u^{\left(0\right)} \in \C^J$
    \item Number of iteration $T$, scaling parameters $\tau$, initial coherence $\rho_0 = 1$
   
    \end{itemize}
     \For {each iteration $i \in [T]$ }
    \State Update $\mbf z$ from \eqref{eq:update_z}
    \State Update  $\bs \theta,\bs \phi,\bs \chi$  from \eqref{eq:update_ang}
    \State Update $\mbf u$ from \eqref{eq:update_u}
      \If {$\underset{ 1 \leq r < q \leq N}{\text{max}} \card{g_{q,r}(\bs \theta^{(i)},\bs \phi^{(i)},\bs \chi^{(i)})} < \rho_{i-1}$ }
    	\State{$\rho_i = \underset{ 1 \leq r < q \leq N}{\text{max}} \card{g_{q,r}(\bs \theta^{(i)},\bs \phi^{(i)},\bs \chi^{(i)})}$}
   	\Else 
   		\State{$\rho_{i} = \rho_{i-1}$}
	\EndIf
    \EndFor   
    \end{algorithmic}
\end{algorithm} 
Equation \eqref{eq:proximal} presents the update procedure for the proximal $\ell_{\infty}$-norm as in  \eqref{eq:update_z}. 
\begin{equation}
 \begin{aligned}
   & \mbf{\hat z}  = \mbf z^{\left(i-1 \right)} - \eta \nabla_{\mbf z } f\left(\theta^{\left(i -1 \right)}, \bs \phi^{\left(i -1 \right)}, \bs \chi^{\left(i -1 \right)}, \mbf z, \mbf u^{\left(i-1\right)}\right)\\
&\mbf z^{\left(i\right)} =  \mbf{\hat z} - \lambda \eta \text{Proj}_{\norm{.}_1}\left( \frac{\mbf{\hat z} }{\eta \lambda} \right)\label{eq:proximal}, 
\end{aligned}   
\end{equation}
where  $\lambda$ is the regularization parameter.
Additionally, we also provide the update for angles $\bs \theta, \bs \phi, \bs \chi$ in \eqref{eq:update_angles} to solve \eqref{eq:update_ang}. 
For simplicity,  we only write the update $\bs \theta$. The update $\bs \phi$ and $\bs \chi$ can be adapted by changing the gradient index.

\begin{equation}
    \begin{aligned}
    \label{eq:update_angles}
     \bs \theta^{\left(i\right)} = \bs\theta^{\left(i-1\right)} - &\eta \nabla_{\bs \theta } f\left(\theta^{\left(i -1 \right)}, \bs \phi^{\left(i -1 \right)}, \bs \chi^{\left(i -1 \right)}, \mbf z^{\left( i\right)}, \mbf u^{\left(i-1\right)}\right) \\
     &\times\nabla_{\bs \theta } \left( g_{q(j),r(j)}(\bs \theta,\bs \phi,\bs \chi) \right)\\
     \end{aligned}
\end{equation}
For both cases, gradient-based and \ac{ALM}-based, the optimization problem for Wigner D-function can be easily tailored for the spherical harmonics case by setting degree $\mu = 0$ in \eqref{coherence_Wigner} \cite{bangun2020dissertation, culotta2021dissertation}.  
\section{Numerical Evaluation}
\label{sec5_NumEval}
In this section, we perform the numerical optimization of the sensing matrix for general construction as discussed in Section \ref{Sec:General}, as well as in Section \ref{Sec:Specific}.
The sampling points from the gradient-based Algorithm \ref{algo_gd}, as well as the \ac{ALM}-based Algorithm \ref{algo_ALM}, are compared to two known sampling patterns: the spiral \cite{saff1997distributing} and the Hammersley \cite{cui1997equidistribution} sampling pattern. For all cases, we do sparse recovery by using YALL1 \cite{yang2011alternating}. The results show the average of $100$ sparse-recovery trials with randomly sparse \ac{SMCs} generated following a zero-mean, unit-variance Gaussian distribution.
\subsection{General Case}
In the general case, the whole range of orders $-n \leq \mu,m \leq n$ is considered. Hence, the sensing matrix is simply given by the expansion of all degrees and orders, as in \eqref{sec3_expand}. For spherical harmonics, the matrix can be derived directly by setting $\mu = 0$ as introduced in \eqref{eq.sh_wign}.
\subsubsection{Wigner D-functions}
\begin{figure}[!htb]
\centering
%
     \scalebox{0.55}{
 
\begin{tikzpicture}

\definecolor{color0}{rgb}{0.580392156862745,0.403921568627451,0.741176470588235}
\definecolor{color1}{rgb}{1,0.498039215686275,0.0549019607843137}
\definecolor{color2}{rgb}{0.549019607843137,0.337254901960784,0.294117647058824}

\begin{axis}[
width=5in,
height=2.5in,
legend columns=4,
legend cell align={left},
legend style={nodes={scale=1.2, transform shape}, fill opacity=0.8, draw opacity=1, text opacity=1, at={(0.5,1.3)}, anchor=north, draw=white!80!black},
tick align=outside,
tick pos=left,
x grid style={white!69.0196078431373!black},
xlabel={\Large Samples (m)},
xmajorgrids,
xmin=17, xmax=81,
xtick style={color=black},
y grid style={white!69.0196078431373!black},
ylabel={\Large Coherence},
ymajorgrids,
ymin=0, ymax=1,
ytick style={color=black}
]
\path [draw=red, fill=red, opacity=0.2]
(axis cs:17,0.563095221045801)
--(axis cs:17,0.447866791594421)
--(axis cs:25,0.334198955286136)
--(axis cs:33,0.275909590385083)
--(axis cs:41,0.228162556732008)
--(axis cs:49,0.195752011333204)
--(axis cs:57,0.16852089937673)
--(axis cs:65,0.154336401127732)
--(axis cs:73,0.134093927474123)
--(axis cs:81,0.122029550608945)
--(axis cs:81,0.150523207488403)
--(axis cs:81,0.150523207488403)
--(axis cs:73,0.164422698291369)
--(axis cs:65,0.184097519911134)
--(axis cs:57,0.200831987722761)
--(axis cs:49,0.242385884298299)
--(axis cs:41,0.272505866386382)
--(axis cs:33,0.319950976422766)
--(axis cs:25,0.400462974443197)
--(axis cs:17,0.563095221045801)
--cycle;

\path [draw=blue, fill=blue, opacity=0.2]
(axis cs:17,0.797158479690552)
--(axis cs:17,0.441577024172618)
--(axis cs:25,0.318205429644922)
--(axis cs:33,0.252950239080231)
--(axis cs:41,0.210355332670615)
--(axis cs:49,0.177256527939974)
--(axis cs:57,0.152612643450448)
--(axis cs:65,0.126820534573274)
--(axis cs:73,0.108464818932745)
--(axis cs:81,0.0901751318123339)
--(axis cs:81,0.10989766303291)
--(axis cs:81,0.10989766303291)
--(axis cs:73,0.13811083137989)
--(axis cs:65,0.160513371229172)
--(axis cs:57,0.196650591201546)
--(axis cs:49,0.221000581979752)
--(axis cs:41,0.265360758180087)
--(axis cs:33,0.309719324111939)
--(axis cs:25,0.413720548152924)
--(axis cs:17,0.797158479690552)
--cycle;

\path [draw=black, fill=black, opacity=0.2]
(axis cs:17,0.731385231018066)
--(axis cs:17,0.402002537821025)
--(axis cs:25,0.300423212888109)
--(axis cs:33,0.240478783845902)
--(axis cs:41,0.204056442230543)
--(axis cs:49,0.17028416595705)
--(axis cs:57,0.148300307683374)
--(axis cs:65,0.130400578520915)
--(axis cs:73,0.120102483035541)
--(axis cs:81,0.10859757010484)
--(axis cs:81,0.117860355291893)
--(axis cs:81,0.117860355291893)
--(axis cs:73,0.129832597469616)
--(axis cs:65,0.146215567173537)
--(axis cs:57,0.16873701410073)
--(axis cs:49,0.19267051819011)
--(axis cs:41,0.231742759358655)
--(axis cs:33,0.281731275109756)
--(axis cs:25,0.397740944900029)
--(axis cs:17,0.731385231018066)
--cycle;

\addplot [line width = 2.5pt, red, solid,mark=*]
table {%
17 0.483780930397865
25 0.36205071746317
33 0.294090289661042
41 0.246102772762435
49 0.212527191398289
57 0.186636795131245
65 0.16558533935252
73 0.147542454066144
81 0.133397803437177
};
\addlegendentry{Sampling in \cite{bangun2021tight}}
\addplot [line width = 2.5pt, blue, solid]
table {%
17 0.545811529919685
25 0.361787275413591
33 0.275238956573269
41 0.224247531106901
49 0.19116875104348
57 0.166136681350527
65 0.13867909232646
73 0.117125459147371
81 0.0993000674119123
};
\addlegendentry{Algorithm \ref{algo_gd}}
\addplot [line width = 2.5pt, black, solid]
table {%
17 0.478009423184972
25 0.325755707447961
33 0.255816232157308
41 0.212628791361711
49 0.1807759197483
57 0.156526269829455
65 0.137282670083628
73 0.12382745962509
81 0.113184291965382
};
\addlegendentry{Algorithm \ref{algo_ALM}}
\addplot [line width = 2.5pt, color0, solid]
table {%
17 0.92517186958175
25 0.948129916821928
33 0.845913408686128
41 0.882378013102918
49 0.861467801809212
57 0.794744160453152
65 0.890971469216527
73 0.889681235086466
81 0.816499614080763
};
\addlegendentry{Spiral \cite{saff1997distributing}}
\addplot [line width = 2.5pt, color1, solid]
table {%
17 0.994955412866983
25 0.997355635131908
33 0.998608803615411
41 0.999036407476908
49 0.999312862226079
57 0.999494254700634
65 0.999625364331356
73 0.999696809856179
81 0.999748707227576
};
\addlegendentry{Hammersley \cite{cui1997equidistribution}}
\addplot [line width = 2.5pt, green!50!black, solid]
table {%
17 0.241384193539774
25 0.170355132992391
33 0.131492840182092
41 0.107024722318207
49 0.0902168216989205
57 0.0779637466033268
65 0.0686370499073717
73 0.0613012037478541
81 0.0553807449914582
};
\addlegendentry{Bound in \cite{bangun_sensing_2020, bangun2021tight}}
\addplot [line width = 2.5pt, color2, solid]
table {%
17 0.217590826036676
25 0.168204289264065
33 0.135931811956759
41 0.111770138901362
49 0.0919887023166835
57 0.074583424470987
65 0.0581129133169545
73 0.0408725415267021
81 0.0173526402098451
};
\addlegendentry{Welchbound \cite{welch_lower_1974}}
\end{axis}

\end{tikzpicture}}  
 
\caption{Coherence of Wigner D-functions sensing matrix}
\label{Fig_Wigner_uniform}
\end{figure}
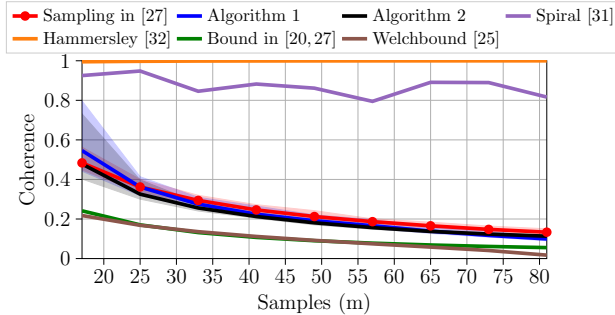

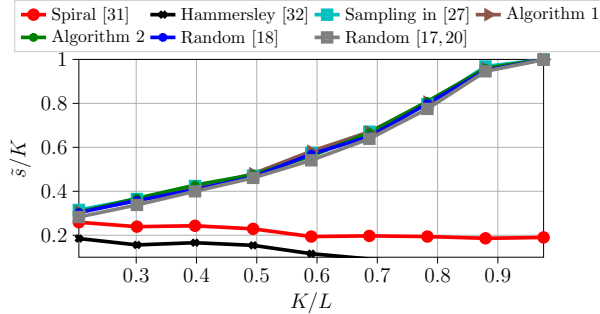
\begin{figure}[!htb]
\centering
%
     \scalebox{0.55}{
\begin{tikzpicture}

\definecolor{color0}{rgb}{0,0.75,0.75}
\definecolor{color1}{rgb}{0.549019607843137,0.337254901960784,0.294117647058824}
\definecolor{color2}{rgb}{0.75,0.75,0}

\begin{axis}[
width=5in,
height=2.5in,
legend cell align={left},
legend columns=4,
legend style={nodes={scale=1.2, transform shape},fill opacity=0.8, draw opacity=1, text opacity=1, at={(0.5,1.3)}, anchor=north, draw=white!80!black},
tick align=outside,
tick pos=left,
x grid style={white!69.0196078431373!black},
xlabel={\Large $K/L$},
xmajorgrids,
xmin=0.204819277108434, xmax=0.975903614457831,
xtick style={color=black},
y grid style={white!69.0196078431373!black},
ylabel={\Large $\tilde s/K$},
ymajorgrids,
ymin=0.1, ymax=1,
ytick style={color=black}
]
\addplot [line width = 2.5pt, red, solid, mark=*, mark size=3, mark options={solid}]
table {%
0.204819277108434 0.259
0.301204819277108 0.239
0.397590361445783 0.243
0.493975903614458 0.229
0.590361445783133 0.194
0.686746987951807 0.197
0.783132530120482 0.194
0.879518072289157 0.186
0.975903614457831 0.19
};
\addlegendentry{Spiral \cite{saff1997distributing}}
\addplot [line width = 2.5pt, black, solid, mark=x, mark size=3, mark options={solid}]
table {%
0.204819277108434 0.185
0.301204819277108 0.156
0.397590361445783 0.166
0.493975903614458 0.154
0.590361445783133 0.116
0.686746987951807 0.092
0.783132530120482 0.076
0.879518072289157 0.066
0.975903614457831 0.051
};
\addlegendentry{Hammersley \cite{cui1997equidistribution}}
\addplot [line width = 2.5pt, color0, solid, mark=square*, mark size=3, mark options={solid}]
table {%
0.204819277108434 0.315
0.301204819277108 0.365
0.397590361445783 0.424
0.493975903614458 0.471
0.590361445783133 0.575
0.686746987951807 0.671
0.783132530120482 0.796
0.879518072289157 0.968
0.975903614457831 1
};
\addlegendentry{Sampling in \cite{bangun2021tight}}
\addplot [line width = 2.5pt, color1, solid, mark=triangle*, mark size=3, mark options={solid,rotate=270}]
table {%
0.204819277108434 0.308
0.301204819277108 0.358
0.397590361445783 0.411
0.493975903614458 0.48
0.590361445783133 0.582
0.686746987951807 0.67
0.783132530120482 0.806
0.879518072289157 0.959
0.975903614457831 1
};
\addlegendentry{Algorithm \ref{algo_gd}}
\addplot [line width = 2.5pt, green!50!black, solid, mark=asterisk, mark size=3, mark options={solid}]
table {%
0.204819277108434 0.305
0.301204819277108 0.368
0.397590361445783 0.428
0.493975903614458 0.476
0.590361445783133 0.56
0.686746987951807 0.666
0.783132530120482 0.809
0.879518072289157 0.959
0.975903614457831 1
};
\addlegendentry{Algorithm \ref{algo_ALM}}
\addplot [line width = 2.5pt, blue, solid, mark=asterisk, mark size=3, mark options={solid}]
table {%
0.204819277108434 0.303
0.301204819277108 0.358
0.397590361445783 0.407
0.493975903614458 0.465
0.590361445783133 0.569
0.686746987951807 0.649
0.783132530120482 0.799
0.879518072289157 0.949
0.975903614457831 1
};
\addlegendentry{Random \cite{burq_weighted_2012}}
\addplot [line width = 2.5pt, white!49.8039215686275!black, solid, mark=square*, mark size=3, mark options={solid}]
table {%
0.204819277108434 0.283
0.301204819277108 0.338
0.397590361445783 0.4
0.493975903614458 0.461
0.590361445783133 0.541
0.686746987951807 0.639
0.783132530120482 0.775
0.879518072289157 0.946
0.975903614457831 0.999
};
\addlegendentry{Random \cite{rauhut_sparse_2011, bangun_sensing_2020}}
\end{axis}

\end{tikzpicture}}  
 
\caption{Phase transition diagram for the sparse recovery with Wigner D-functions, general case.}
\label{Fig_pt_wigner}
\end{figure}

The numerical coherence for a sensing matrix from Wigner D-functions when all angles are optimized is given in Fig.~\ref{Fig_Wigner_uniform}, where in this case, we evaluate $N = 3$ with column dimension $L = 83$. Since our algorithms are initialized randomly, the shown coherence is an average of the results, presented together with their deviation.
For the spiral and Hammersley sampling schemes, the samples along the polarization angle are distributed evenly within the interval $\chi \in [0,2\pi)$. It can be seen that these sampling points generate the worst coherence compared to the sampling points produced by considering the gradient method and \ac{ALM}, described in Algorithm \ref{algo_gd} and \ref{algo_ALM}. Furthermore, the performance of optimized sensing matrices in terms of sparse recovery by using a phase transition diagram is presented in Fig.~\ref{Fig_pt_wigner} . In contrast to the spiral and Hammersley samplings, that yield a high coherence sensing matrix, the optimized sampling points using the gradient descent and \ac{ALM} deliver a better recovery and performance similar to the random sampling's. The graphs depict the transition bounds at a $50\%$ success of recovery with \ac{BP} as in \eqref{ch2:eq P1}. The bounds are tight, so that probability of success is $100\%$ almost immediately below the curve, while it is $0\%$ almost immediately above the curve. The abscissa shows $(K/L)$, that is, the number of samples $K$ with respect to the total number of unknowns $L$, while the ordinate shows the ratio between the sparsity of the vector $\tilde{s}$ and the number of samples acquired $K$. 
\subsubsection{Spherical harmonics}
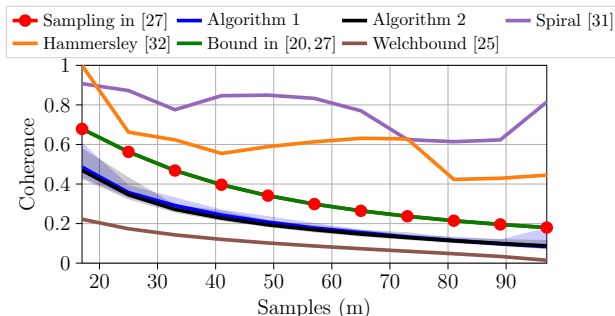
\begin{figure}[!htb]
\centering
%
     \scalebox{0.55}{
\begin{tikzpicture}

\definecolor{color0}{rgb}{0.580392156862745,0.403921568627451,0.741176470588235}
\definecolor{color1}{rgb}{1,0.498039215686275,0.0549019607843137}
\definecolor{color2}{rgb}{0.549019607843137,0.337254901960784,0.294117647058824}

\begin{axis}[
width=5in,
height=2.5in,
legend columns=4,
legend cell align={left},
legend style={nodes={scale=1.2, transform shape},fill opacity=0.8, draw opacity=1, text opacity=1, at={(0.5,1.3)}, anchor = north, draw=white!80!black},
tick align=outside,
tick pos=left,
x grid style={white!69.0196078431373!black},
xlabel={\Large Samples (m)},
xmajorgrids,
xmin=17, xmax=97,
xtick style={color=black},
y grid style={white!69.0196078431373!black},
ylabel={\Large Coherence},
ymajorgrids,
ymin=0, ymax=1,
ytick style={color=black}
]
\path [draw=red, fill=red, opacity=0.2]
(axis cs:17,0.678688585148102)
--(axis cs:17,0.678688585148102)
--(axis cs:25,0.562539483279067)
--(axis cs:33,0.468247493229728)
--(axis cs:41,0.396366496625075)
--(axis cs:49,0.34134648342864)
--(axis cs:57,0.29852454235762)
--(axis cs:65,0.264555317970759)
--(axis cs:73,0.237109582915589)
--(axis cs:81,0.214560787501565)
--(axis cs:89,0.195757082980803)
--(axis cs:97,0.179868351875199)
--(axis cs:97,0.179868351875199)
--(axis cs:97,0.179868351875199)
--(axis cs:89,0.195757082980803)
--(axis cs:81,0.214561165366173)
--(axis cs:73,0.237109582915589)
--(axis cs:65,0.264555317970759)
--(axis cs:57,0.29852454235762)
--(axis cs:49,0.34134648342864)
--(axis cs:41,0.396366496625075)
--(axis cs:33,0.468247493229728)
--(axis cs:25,0.562539483279067)
--(axis cs:17,0.678688585148102)
--cycle;

\path [draw=blue, fill=blue, opacity=0.2]
(axis cs:17,0.60583754371336)
--(axis cs:17,0.435898532401022)
--(axis cs:25,0.334152514330376)
--(axis cs:33,0.269833710612692)
--(axis cs:41,0.226196447017496)
--(axis cs:49,0.191441322401206)
--(axis cs:57,0.165949105501978)
--(axis cs:65,0.141253690533984)
--(axis cs:73,0.125061320370803)
--(axis cs:81,0.105129680885704)
--(axis cs:89,0.0892518470199091)
--(axis cs:97,0.0725258329614922)
--(axis cs:97,0.176065521487288)
--(axis cs:97,0.176065521487288)
--(axis cs:89,0.117835309522151)
--(axis cs:81,0.131365623354751)
--(axis cs:73,0.141183949473325)
--(axis cs:65,0.167412451414938)
--(axis cs:57,0.200231460704922)
--(axis cs:49,0.235073500639986)
--(axis cs:41,0.269287289367543)
--(axis cs:33,0.329735333934972)
--(axis cs:25,0.391600855056434)
--(axis cs:17,0.60583754371336)
--cycle;

\path [draw=black, fill=black, opacity=0.2]
(axis cs:17,0.578137218425572)
--(axis cs:17,0.428248160472076)
--(axis cs:25,0.324725731343097)
--(axis cs:33,0.25751069089351)
--(axis cs:41,0.216228664945521)
--(axis cs:49,0.184485982841048)
--(axis cs:57,0.157717333791767)
--(axis cs:65,0.136545429784921)
--(axis cs:73,0.1196843147737)
--(axis cs:81,0.103079947305742)
--(axis cs:89,0.0906551390550579)
--(axis cs:97,0.0781537700688023)
--(axis cs:97,0.114240678486142)
--(axis cs:97,0.114240678486142)
--(axis cs:89,0.119188697196775)
--(axis cs:81,0.126264553653844)
--(axis cs:73,0.153586315068298)
--(axis cs:65,0.166516279080309)
--(axis cs:57,0.189267222905105)
--(axis cs:49,0.213980135456048)
--(axis cs:41,0.263364519138523)
--(axis cs:33,0.292927907556372)
--(axis cs:25,0.431878457855698)
--(axis cs:17,0.578137218425572)
--cycle;

\addplot [line width = 2.5pt, red, solid, mark=*, mark size=3, mark options={solid}]
table {%
17 0.678688585148102
25 0.562539483279067
33 0.468247493229728
41 0.396366496625075
49 0.34134648342864
57 0.29852454235762
65 0.264555317970759
73 0.237109582915589
81 0.214560793799309
89 0.195757082980803
97 0.179868351875199
};
\addlegendentry{Sampling in \cite{bangun2021tight}}
\addplot [line width = 2.5pt, blue, solid]
table {%
17 0.485565221177662
25 0.356802541110804
33 0.291130956896944
41 0.242993778174463
49 0.20720088078067
57 0.17749764978224
65 0.154182163888756
73 0.132715181719764
81 0.114254495727893
89 0.0976012824010995
97 0.0833823904923566
};
\addlegendentry{Algorithm \ref{algo_gd}}
\addplot [line width = 2.5pt, black, solid]
table {%
17 0.469105784887062
25 0.346066579042004
33 0.271970268574155
41 0.227745089954814
49 0.193733530151868
57 0.168805224390376
65 0.147283365033955
73 0.129383685352059
81 0.113365667713165
89 0.0996096709208727
97 0.0874158994796251
};
\addlegendentry{Algorithm \ref{algo_ALM}}
\addplot [line width = 2.5pt, color0, solid]
table {%
17 0.907308888712137
25 0.872404065864269
33 0.775705831840134
41 0.846473863499718
49 0.849394784754606
57 0.832999468507669
65 0.770444785724877
73 0.62468373907765
81 0.614053929946316
89 0.623086140164811
97 0.814541809353065
};
\addlegendentry{Spiral \cite{saff1997distributing}}
\addplot [line width = 2.5pt, color1, solid]
table {%
17 0.999350946544344
25 0.662225087111935
33 0.623270075336833
41 0.554604371485608
49 0.588924361323435
57 0.613322160907507
65 0.631240492096142
73 0.628206712815119
81 0.42319211023221
89 0.42907055189093
97 0.445046415599537
};
\addlegendentry{Hammersley \cite{cui1997equidistribution}}
\addplot [line width = 2.5pt, green!50!black, solid]
table {%
17 0.678688585148103
25 0.562539483279067
33 0.468247493229728
41 0.396366496625075
49 0.34134648342864
57 0.298524542357621
65 0.264555317970759
73 0.237109582915589
81 0.214560787501566
89 0.195757082980803
97 0.179868351875199
};
\addlegendentry{Bound in \cite{bangun_sensing_2020, bangun2021tight}}
\addplot [line width = 2.5pt, color2, solid]
table {%
17 0.221855105587296
25 0.173793215151378
33 0.142857142857143
41 0.120145920902904
49 0.102040816326531
57 0.086710996952412
65 0.073058330279537
73 0.060285408222111
81 0.0476190476190476
89 0.0338603902242841
97 0.0145049452161946
};
\addlegendentry{Welchbound \cite{welch_lower_1974}}
\end{axis}

\end{tikzpicture}}  
 
\caption{Coherence of sensing matrices from spherical harmonics.}
\label{Fig_SH_uniform}
\end{figure}

 The main difference with the Wigner D case is the lack of polarization angle $\chi \in [0,2\pi)$. Thereby, we only optimize the elevation $\theta$ and azimuth $\phi$ angles. The average coherence of a sensing matrix constructed with spherical harmonics with $N = 9$ and column dimension of the matrix $L = 99$ is shown in Fig.~\ref{Fig_SH_uniform}.
Similar to the results shown for Wigner D-functions, the optimized sampling points yield a lower-coherence sensing matrix for the spherical harmonics case compared to the construction from spiral and Hammersley schemes. Additionally, optimizing the sampling both on elevation $\theta$ and azimuth $\phi$ yields a lower coherence compared to the theoretical bound, as discussed in \cite{bangun_sensing_2020, bangun2021tight}, at least numerically. 

\begin{figure}[!htb]
\centering
\includegraphics[scale=0.45]{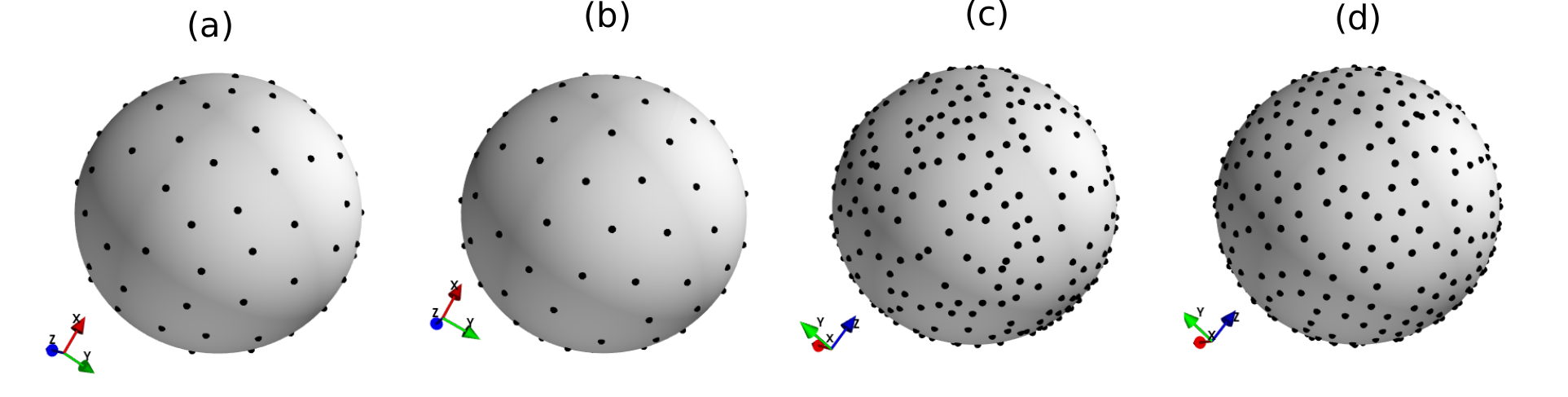}
\caption{Sampling distribution from an optimized sensing matrix from spherical harmonics. (a) \ac{ALM} $\left(K = 97 , N = 9 \right)$ (b) Grad. Descent $\left(K =  97  , N = 9 \right)$ (c) $\ac{ALM} \left(K =  400 , N =  21  \right)$ (d) Grad. Descent $\left(K =  400 , N =  21 \right)$}
\label{fig.sphere_dist}
\end{figure}

Fig.~\ref{fig.sphere_dist} shows the distribution of the optimized sampling points on the sphere for the spherical harmonics case. The distribution follows a specific pattern that is approximately equidistant on a geodesic, which demonstrates uniformity of the optimized sampling points.

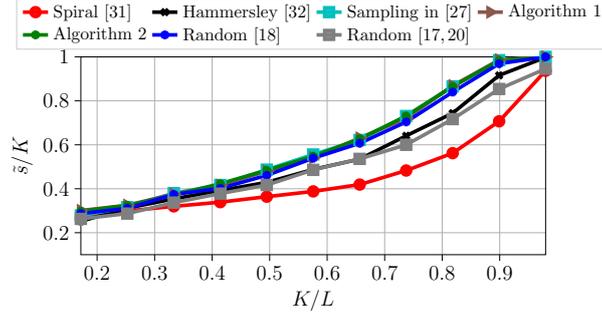
\begin{figure}[!htb]
\centering
%
     \scalebox{0.55}{
\begin{tikzpicture}

\definecolor{color0}{rgb}{0,0.75,0.75}
\definecolor{color1}{rgb}{0.549019607843137,0.337254901960784,0.294117647058824}
\definecolor{color2}{rgb}{0.75,0.75,0}

\begin{axis}[
width=5in,
height=2.5in,
legend columns=4,
legend cell align={left},
legend style={nodes={scale=1.2, transform shape}, fill opacity=0.8, draw opacity=1, text opacity=1,  at={(0.5,1.3)}, anchor=north, draw=white!80!black},
tick align=outside,
tick pos=left,
x grid style={white!69.0196078431373!black},
xlabel={\Large $K/L$},
xmajorgrids,
xmin=0.171717171717172, xmax=0.97979797979798,
xtick style={color=black},
y grid style={white!69.0196078431373!black},
ylabel={\Large $\tilde s/K$},
ymajorgrids,
ymin=0.1, ymax=1,
ytick style={color=black}
]
\addplot [line width = 2.5pt, red, solid, mark=*, mark size=3, mark options={solid}]
table {%
0.171717171717172 0.265
0.252525252525253 0.299
0.333333333333333 0.32
0.414141414141414 0.339
0.494949494949495 0.364
0.575757575757576 0.388
0.656565656565657 0.419
0.737373737373737 0.483
0.818181818181818 0.562
0.898989898989899 0.707
0.97979797979798 0.937
};
\addlegendentry{Spiral \cite{saff1997distributing}}
\addplot [line width = 2.5pt, black, solid, mark=x, mark size=3, mark options={solid}]
table {%
0.171717171717172 0.252
0.252525252525253 0.312
0.333333333333333 0.354
0.414141414141414 0.392
0.494949494949495 0.429
0.575757575757576 0.488
0.656565656565657 0.535
0.737373737373737 0.641
0.818181818181818 0.744
0.898989898989899 0.916
0.97979797979798 0.998
};
\addlegendentry{Hammersley \cite{cui1997equidistribution}}
\addplot [line width = 2.5pt, color0, solid, mark=square*, mark size=3, mark options={solid}]
table {%
0.171717171717172 0.278
0.252525252525253 0.316
0.333333333333333 0.38
0.414141414141414 0.417
0.494949494949495 0.487
0.575757575757576 0.556
0.656565656565657 0.622
0.737373737373737 0.731
0.818181818181818 0.866
0.898989898989899 0.985
0.97979797979798 1
};
\addlegendentry{Sampling in \cite{bangun2021tight}}
\addplot [line width = 2.5pt, color1, solid, mark=triangle*, mark size=3, mark options={solid,rotate=270}]
table {%
0.171717171717172 0.301
0.252525252525253 0.325
0.333333333333333 0.376
0.414141414141414 0.416
0.494949494949495 0.48
0.575757575757576 0.541
0.656565656565657 0.631
0.737373737373737 0.72
0.818181818181818 0.869
0.898989898989899 0.986
0.97979797979798 1
};
\addlegendentry{Algorithm \ref{algo_gd}}
\addplot [line width = 2.5pt, green!50!black, solid, mark=asterisk, mark size=3, mark options={solid}]
table {%
0.171717171717172 0.298
0.252525252525253 0.324
0.333333333333333 0.371
0.414141414141414 0.423
0.494949494949495 0.484
0.575757575757576 0.548
0.656565656565657 0.627
0.737373737373737 0.731
0.818181818181818 0.865
0.898989898989899 0.983
0.97979797979798 1
};
\addlegendentry{Algorithm \ref{algo_ALM}}
\addplot [line width = 2.5pt, blue, solid, mark=asterisk, mark size=3, mark options={solid}]
table {%
0.171717171717172 0.287
0.252525252525253 0.311
0.333333333333333 0.374
0.414141414141414 0.402
0.494949494949495 0.461
0.575757575757576 0.539
0.656565656565657 0.607
0.737373737373737 0.704
0.818181818181818 0.84
0.898989898989899 0.969
0.97979797979798 1
};
\addlegendentry{Random \cite{burq_weighted_2012}}
\addplot [line width = 2.5pt, white!49.8039215686275!black, solid, mark=square*, mark size=3, mark options={solid}]
table {%
0.171717171717172 0.263
0.252525252525253 0.287
0.333333333333333 0.337
0.414141414141414 0.378
0.494949494949495 0.417
0.575757575757576 0.486
0.656565656565657 0.536
0.737373737373737 0.6
0.818181818181818 0.717
0.898989898989899 0.854
0.97979797979798 0.946
};
\addlegendentry{Random \cite{rauhut_sparse_2011, bangun_sensing_2020}}
\end{axis}

\end{tikzpicture}}  
 
\caption{Phase transition diagram for sparse recovery of spherical harmonics.}
\label{Fig_pt_sh}
\end{figure}

The phase transition diagram is depicted in Fig.~\ref{Fig_pt_sh}. It can be seen that the optimized sensing matrix yields a performance similar to the random sampling's, and shows a better recovery compared to the spiral and Hammersley samplings. 
\subsection{Specific Case: Practical SNF Measurements}
In most cases, we only evaluate the order $\mu = \pm 1$ with linearly polarized probes for polarization angles $\chi = 0$ and $\chi = \frac{\pi}{2}$. This condition is not contrived and has practical relevance, as discussed in \cite{hansen1988spherical, wacker1975non}. In addition, we can provide experimental data from real measurements, which makes this case even more interesting, as explained in Section \ref{Sec:Specific}. Additionally, in this section we perform the far-field reconstruction with fewer sampling points, i.e., $K < L$, compared to the conventional settings that require a total of $K > 2L$ measurements.
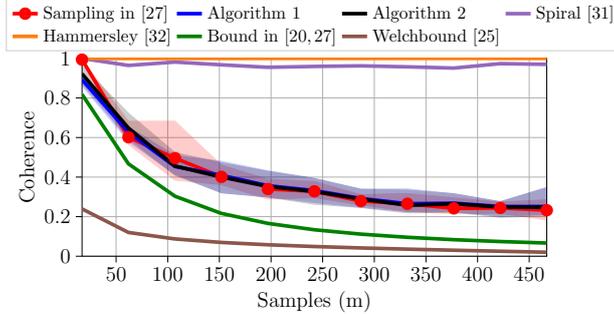
\begin{figure}[!htb]
\centering
%
     \scalebox{0.55}{
\begin{tikzpicture}

\definecolor{color0}{rgb}{0.580392156862745,0.403921568627451,0.741176470588235}
\definecolor{color1}{rgb}{1,0.498039215686275,0.0549019607843137}
\definecolor{color2}{rgb}{0.549019607843137,0.337254901960784,0.294117647058824}

\begin{axis}[
width=5in,
height=2.5in,
legend columns=4,
legend cell align={left},
legend style={nodes={scale=1.2, transform shape}, fill opacity=0.8, draw opacity=1, text opacity=1, at={(0.5,1.3)}, anchor=north, draw=white!80!black},
tick align=outside,
tick pos=left,
x grid style={white!69.0196078431373!black},
xlabel={\Large Samples (m)},
xmajorgrids,
xmin=17, xmax=467,
xtick style={color=black},
y grid style={white!69.0196078431373!black},
ylabel={\Large Coherence},
ymajorgrids,
ymin=0, ymax=1,
ytick style={color=black}
]
\path [draw=red, fill=red, opacity=0.2]
(axis cs:17,0.993936717510223)
--(axis cs:17,0.993936717510223)
--(axis cs:62,0.565562387129653)
--(axis cs:107,0.384903971817785)
--(axis cs:152,0.361820383273184)
--(axis cs:197,0.288721222531841)
--(axis cs:242,0.296506802339006)
--(axis cs:287,0.259593589593829)
--(axis cs:332,0.226500048937063)
--(axis cs:377,0.214358844932108)
--(axis cs:422,0.217325486821278)
--(axis cs:467,0.182975619780603)
--(axis cs:467,0.28582676939909)
--(axis cs:467,0.28582676939909)
--(axis cs:422,0.273388365499599)
--(axis cs:377,0.29499719735611)
--(axis cs:332,0.316675213910571)
--(axis cs:287,0.311005135544096)
--(axis cs:242,0.377341217897012)
--(axis cs:197,0.389142749809636)
--(axis cs:152,0.460521706441354)
--(axis cs:107,0.682788492548722)
--(axis cs:62,0.681922229925254)
--(axis cs:17,0.993936717510223)
--cycle;

\path [draw=blue, fill=blue, opacity=0.2]
(axis cs:17,0.926958722803987)
--(axis cs:17,0.86111319065094)
--(axis cs:62,0.580610394477845)
--(axis cs:107,0.414223131252723)
--(axis cs:152,0.320784476747405)
--(axis cs:197,0.299268282232029)
--(axis cs:242,0.272108435630798)
--(axis cs:287,0.247374832630157)
--(axis cs:332,0.225027962120496)
--(axis cs:377,0.225222163217468)
--(axis cs:422,0.202233098365738)
--(axis cs:467,0.211578119617016)
--(axis cs:467,0.348365169236758)
--(axis cs:467,0.348365169236758)
--(axis cs:422,0.275637809333213)
--(axis cs:377,0.315972626209259)
--(axis cs:332,0.340535367621439)
--(axis cs:287,0.338886732660262)
--(axis cs:242,0.393261726908834)
--(axis cs:197,0.431747265507342)
--(axis cs:152,0.479887932538986)
--(axis cs:107,0.508958599083662)
--(axis cs:62,0.653988003730774)
--(axis cs:17,0.926958722803987)
--cycle;

\path [draw=black, fill=black, opacity=0.2]
(axis cs:17,0.963277995586395)
--(axis cs:17,0.872616410255432)
--(axis cs:62,0.593604445457458)
--(axis cs:107,0.409769870642475)
--(axis cs:152,0.32068810022756)
--(axis cs:197,0.299220389968969)
--(axis cs:242,0.262861306958765)
--(axis cs:287,0.246332705020905)
--(axis cs:332,0.219001464878317)
--(axis cs:377,0.222854964794339)
--(axis cs:422,0.196872681379318)
--(axis cs:467,0.204908250298814)
--(axis cs:467,0.345883848451458)
--(axis cs:467,0.345883848451458)
--(axis cs:422,0.275989592075348)
--(axis cs:377,0.315972626209259)
--(axis cs:332,0.33397247410671)
--(axis cs:287,0.338994894010484)
--(axis cs:242,0.393261726908834)
--(axis cs:197,0.429012963315213)
--(axis cs:152,0.472623783762936)
--(axis cs:107,0.52399369032148)
--(axis cs:62,0.72272156503298)
--(axis cs:17,0.963277995586395)
--cycle;

\addplot [line width = 2.5pt, red, solid, mark=*, mark size=3, mark options={solid}]
table {%
17 0.993936717510223
62 0.601511193161579
107 0.495631717935185
152 0.401216536115375

197 0.339793246166516
242 0.327966121157496
287 0.278658699184112
332 0.265383561136868
377 0.24207638044146
422 0.243852860905558
467 0.232606128023376
};
\addlegendentry{Sampling in \cite{bangun2021tight}}
\addplot [line width = 2.5pt, blue, solid]
table {%
17 0.892422078420019
62 0.626635272678183
107 0.453671532603623
152 0.408963948889319
197 0.356505786016393
242 0.331750965641877
287 0.292157958910769
332 0.26534871846323
377 0.26802824355382
422 0.25126003886689
467 0.251427434431673
};
\addlegendentry{Algorithm \ref{algo_gd}}
\addplot [line width = 2.5pt, black, solid]
table {%
17 0.9221514105221
62 0.647428392669959
107 0.455815140429808
152 0.400321584482742
197 0.352539633920831
242 0.327012131027862
287 0.289372701686201
332 0.257532147755335
377 0.264925951923443
422 0.248442054403113
467 0.247698681393055
};
\addlegendentry{Algorithm \ref{algo_ALM}}
\addplot [line width = 2.5pt, color0, solid]
table {%
17 1.00000011920929
62 0.964613926939821
107 0.981603416571636
152 0.967597545322357
197 0.955084469535071
242 0.959488245193862
287 0.962448004002887
332 0.957441590613299
377 0.951155124804006
422 0.973623615807197
467 0.969958769761094
};
\addlegendentry{Spiral \cite{saff1997distributing}}
\addplot [line width = 2.5pt, color1, solid]
table {%
17 1.00000011920929
62 0.999686459352468
107 0.999880387689126
152 0.99993735528187
197 0.999960841107565
242 0.999973673804372
287 0.999981164933747
332 0.999985934421613
377 0.999989331643386
422 0.999991422535412
467 0.999993089875604
};
\addlegendentry{Hammersley \cite{cui1997equidistribution}}
\addplot [line width = 2.5pt, green!50!black, solid]
table {%
17 0.818027220803633
62 0.46686354886922
107 0.303370933602523
152 0.216323887649468
197 0.165837990646581
242 0.133656752917883
287 0.111596337705257
332 0.0956240816781205
377 0.0835660669060788
422 0.0741602731819398
467 0.0666287344272294
};
\addlegendentry{Bound in \cite{bangun_sensing_2020, bangun2021tight}}
\addplot [line width = 2.5pt, color2, solid]
table {%
17 0.239137409802949
62 0.120074778010061
107 0.0873093208838897
152 0.0696509768864421
197 0.0578432344399074
242 0.0489927286398298
287 0.0418479367548121
332 0.0357513091043049
377 0.0302985566849799
422 0.0251924388226035
467 0.0201474870984932
};
\addlegendentry{Welchbound \cite{welch_lower_1974}}
\end{axis}

\end{tikzpicture}}  
 
\caption{Coherence of sensing matrices for practical \ac{SNF} measurements.}
\label{Fig_snf_uniform}
\end{figure} 

Since we only consider $\mu = \pm 1$ and polarization angles $\chi = 0$ and $\chi = \frac{\pi}{2}$, we only optimize the elevation $\theta $ and azimuth $\phi$ angles. Fig.~\ref{Fig_snf_uniform} presents the coherence of the sensing matrix from a specific setting constructed from several sampling points. The difference between this specific case compared to the general case in the previous section is that our sensing matrix results from concatenating the matrix constructed from Wigner D-functions for both values of $\chi$ considered, as derived in \eqref{sec3_expand2}. The optimized sensing matrix, again, produces lower coherence compared to known deterministic sampling schemes. 

\begin{figure}[!htb]
\centering
%
     \scalebox{0.6}{
\begin{tikzpicture}

\definecolor{color0}{rgb}{0,0.75,0.75}
\definecolor{color1}{rgb}{0.549019607843137,0.337254901960784,0.294117647058824}
\definecolor{color2}{rgb}{0.75,0.75,0}

\begin{axis}[
width=5in,
height=2.5in,
legend columns=4,
legend cell align={left},
legend style={nodes={scale=1.2, transform shape}, fill opacity=0.8, draw opacity=1, text opacity=1,  at={(0.5,1.3)}, anchor=north, draw=white!80!black},
tick align=outside,
tick pos=left,
x grid style={white!69.0196078431373!black},
xlabel style={font=\color{white!15!black}},
xlabel={\Large $K/L$},
xmajorgrids,
xmin=0.0295138888888889, xmax=0.810763888888889,
xtick style={color=black},
y grid style={white!69.0196078431373!black},
ylabel style={font=\color{white!15!black}},
ylabel={\Large $\tilde s/K$},
ymajorgrids,
ymin=0.1, ymax=1,
ytick style={color=black}
]
\addplot [line width = 2.5pt, red, solid, mark=*, mark size=3, mark options={solid}]
table {%
0.0295138888888889 0.004
0.107638888888889 0.211
0.185763888888889 0.264
0.263888888888889 0.308
0.342013888888889 0.347
0.420138888888889 0.384
0.498263888888889 0.439
0.576388888888889 0.5
0.654513888888889 0.554
0.732638888888889 0.621
0.810763888888889 0.724
};
\addlegendentry{Spiral \cite{saff1997distributing}}
\addplot [line width = 2.5pt, black, solid, mark=x, mark size=3, mark options={solid}]
table {%
0.0295138888888889 0.005
0.107638888888889 0.18
0.185763888888889 0.229
0.263888888888889 0.231
0.342013888888889 0.235
0.420138888888889 0.18
0.498263888888889 0.177
0.576388888888889 0.167
0.654513888888889 0.169
0.732638888888889 0.185
0.810763888888889 0.17
};
\addlegendentry{Hammersley \cite{cui1997equidistribution}}
\addplot [line width = 2.5pt, color0, solid, mark=square*, mark size=3, mark options={solid}]
table {%
0.0295138888888889 0.008
0.107638888888889 0.213
0.185763888888889 0.272
0.263888888888889 0.308
0.342013888888889 0.363
0.420138888888889 0.411
0.498263888888889 0.462
0.576388888888889 0.522
0.654513888888889 0.595
0.732638888888889 0.704
0.810763888888889 0.832
};
\addlegendentry{Sampling in \cite{bangun2021tight}}
\addplot [line width = 2.5pt, color1, solid, mark=triangle*, mark size=3, mark options={solid,rotate=270}]
table {%
0.0295138888888889 0.009
0.107638888888889 0.207
0.185763888888889 0.28
0.263888888888889 0.308
0.342013888888889 0.364
0.420138888888889 0.409
0.498263888888889 0.466
0.576388888888889 0.523
0.654513888888889 0.595
0.732638888888889 0.696
0.810763888888889 0.821
};
\addlegendentry{Algorithm \ref{algo_gd}}
\addplot [line width = 2.5pt, green!50!black, solid, mark=asterisk, mark size=3, mark options={solid}]
table {%
0.0295138888888889 0.003
0.107638888888889 0.209
0.185763888888889 0.279
0.263888888888889 0.307
0.342013888888889 0.364
0.420138888888889 0.409
0.498263888888889 0.466
0.576388888888889 0.526
0.654513888888889 0.593
0.732638888888889 0.698
0.810763888888889 0.825
};
\addlegendentry{Algorithm \ref{algo_ALM}}
\addplot [line width = 2.5pt, blue, solid, mark=asterisk, mark size=3, mark options={solid}]
table {%
0.0295138888888889 0.043
0.107638888888889 0.215
0.185763888888889 0.275
0.263888888888889 0.31
0.342013888888889 0.368
0.420138888888889 0.402
0.498263888888889 0.465
0.576388888888889 0.519
0.654513888888889 0.595
0.732638888888889 0.692
0.810763888888889 0.821
};
\addlegendentry{Random \cite{burq_weighted_2012}}
\addplot [line width = 2.5pt, white!49.8039215686275!black, solid, mark=square*, mark size=3, mark options={solid}]
table {%
0.0295138888888889 0.04
0.107638888888889 0.202
0.185763888888889 0.247
0.263888888888889 0.291
0.342013888888889 0.326
0.420138888888889 0.371
0.498263888888889 0.404
0.576388888888889 0.461
0.654513888888889 0.502
0.732638888888889 0.56
0.810763888888889 0.637
};
\addlegendentry{Random \cite{rauhut_sparse_2011, bangun_sensing_2020}}
\end{axis}

\end{tikzpicture}}  
 
\caption{Phase transition diagram for sparse recovery for practical \ac{SNF} measurements.}
\label{Fig_pt_snf}
\end{figure}
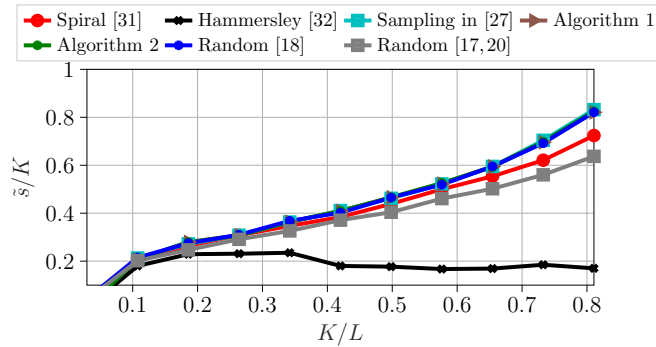  

Furthermore, the optimized sensing matrix also shows higher recovery chances than the spiral and Hammersley sampling schemes, as depicted in Fig.~\ref{Fig_pt_snf}. Moreover, compared to random sampling points with arbitrary polarization angles $\chi$, the optimized sampling points deliver a similar performance while keeping the polarization angles alternate between $\chi = 0$ and $\chi = \frac{\pi}{2}$.

\section{Conclusions}\label{sec7_conclusion}
We have discussed the methods to optimize sensing matrices for the \ac{CS}-based \ac{SNF} method. In contrast to the conventional method, which requires an oversampled near-field measurement, the \ac{CS}-based method can significantly reduce the number of samples to reconstruct the far-field pattern by using an optimized sensing matrix. The strategy emanates from the fact that, in most cases, the \ac{SMCs} are compressible. Harnessing this information is key in implementing the \ac{CS}-based method.

We propose two different approaches, namely gradient-based and \ac{ALM} based proximal methods, to design low coherence sensing matrices that are suitable for the \ac{CS}-based method. Numerical experiments show that these approaches are successful in minimizing the coherence of sensing matrices constructed from Wigner D-functions and spherical harmonics. Numerical simulations from measurement data show that the proposed sampling schemes can reconstruct the far-field pattern with a reasonable error while reducing the number of sampling points. Additionally, numerical experiments show that the optimization methods maintain the uniformity structure of proposed sampling points, hence making it easier to implement in real measurement system.

\section*{Notation} \label{Sect:Appendix}
The vectors and matrices are denoted by bold small-cap letters $\mathbf{a}$ and big-cap letters $\mbf A$. 
The elevation, azimuth, and polarization angles are denoted by $\theta \in [0, \pi]$, $\phi \in [0, 2\pi)$, and $\chi \in [0, 2\pi)$, respectively. The set $\{1, ..., m\}$ is denoted by $[m]$. $\overline{x}$ is the conjugate of $x$. For a vector $\mathbf{x} \in \C^L$, the $\ell_p$-norm is given by $\norm{\mathbf{x}}_p = \left(\sum_{i = 1}^L \card{x_i}^p \right)^{1/p}$ for $p \in [1,\infty)$ and $\norm{\mathbf{x}}_{\infty} = \underset{i \in [L]}{\text{max}} \card{x_i}$ for $p = \infty$. 



\bibliographystyle{IEEEtran}
\bibliography{reference}

\end{document}